\documentclass[10pt]{article}

\usepackage[margin=1in]{geometry}
\usepackage{booktabs}
\usepackage{caption}
\usepackage{enumitem}
\usepackage[hidelinks,pdftitle={When the Approved File Is Unchanged: Authority-Set Integrity for Deployment and Coding-Agent Approvals},pdfauthor={Md Imran Hossen}]{hyperref}
\usepackage{xcolor}
\usepackage{amsmath}
\usepackage{cite}
\usepackage{array}
\usepackage{listings}
\usepackage{tikz}
\usetikzlibrary{arrows.meta,positioning,fit,calc}
\usepackage{pgfplots}
\pgfplotsset{compat=1.16}
\newcolumntype{L}[1]{>{\raggedright\arraybackslash}p{#1}}
\newcommand{\sha}{\mathsf{H}}
\definecolor{accent}{RGB}{178,34,34}
\title{When the Approved File Is Unchanged: Authority-Set Integrity for Deployment and Coding-Agent Approvals}
\author{Md Imran Hossen\\
Department of Computer Science and Engineering\\
University of Tennessee at Chattanooga\\
\texttt{mdimran-hossen@utc.edu}}
\date{}

\begin{document}
\maketitle

\begin{abstract}
An approval for a deployment, a release, or an agent-issued command is recorded against one object: a workflow file, a script, a command line. The run that follows also reads other repository files, which decide what the deployment creates and which changes a policy allows, so an edit to one of them after approval passes any check on the approved object. Such edits are the common case. In the history of 20 public repositories, 296 of the 542 commits that change a file the run reads (54.6\%) leave the approved file untouched, all of them ordinary development with no adversary involved. We demonstrate the consequence on two of those repositories: an edit outside the approved workflow adds a bucket-write policy to a deployed role, a one-line edit removes the human approval required for high-risk changes, and a check on the approved file accepts both. We define \emph{authority-set integrity}: an approval is sound when it binds every repository file the authorized run is configured to read, and exact when it binds no other file. A binding over that set rejects both edits and re-requests a standing approval on about one commit in ten. The hard part is naming the set. Coding agents asked to produce the binding compute the digest correctly and bind different file sets from run to run, so a checker cannot take the set on the agent's word. It should compute the digest itself and draw the set from a source it can verify, such as the deployment tool's own record of its inputs or a manifest declared in the repository.
\end{abstract}

\section{Introduction}

Software delivery places an approval in front of consequential operations. A job that references a protected deployment environment does not proceed until a required reviewer approves it~\cite{github_environments}. A coding agent asks the user before it executes a command. A policy bot records that a change was reviewed. In each case the decision is recorded against an object, and a checker later compares that record with what is about to run.

\begin{figure*}[t]
\centering
\begin{tikzpicture}[x=1mm, y=1mm, font=\sffamily\scriptsize, >={Latex[length=1.6mm,width=1.3mm]},
  panel/.style={fill=black!3, draw=black!12, rounded corners=1.5pt},
  stepno/.style={circle, fill=black!80, text=white, font=\sffamily\bfseries\scriptsize, inner sep=0pt, minimum size=4mm},
  head/.style={font=\sffamily\bfseries\footnotesize, anchor=west, text=black!90},
  sub/.style={font=\sffamily\scriptsize, anchor=west, text=black!60},
  card/.style={draw=black!55, fill=white, rounded corners=0.8pt, minimum height=5.6mm, minimum width=41mm, inner xsep=2mm, anchor=north west, font=\ttfamily\scriptsize, text=black, line width=0.4pt},
  tree/.style={draw=black!45, line width=0.5pt},
  via/.style={font=\sffamily\itshape\scriptsize, text=black!60, anchor=west, inner sep=0.6pt},
  note/.style={font=\sffamily\scriptsize, text=black!70, align=flush left, anchor=north west, inner sep=0pt},
  badge/.style={font=\sffamily\bfseries\scriptsize, text=white, rounded corners=0.8pt, inner xsep=1.6mm, inner ysep=0.9mm, anchor=north east},
  verdict/.style={draw=black!45, fill=white, rounded corners=1pt, line width=0.4pt}]

\fill[panel] (0,0) rectangle (54,-45);
\fill[panel] (57,0) rectangle (111,-45);
\fill[panel] (114,0) rectangle (165,-45);
\draw[->, black!40, line width=0.6pt] (54,-22.5) -- (57,-22.5);
\draw[->, black!40, line width=0.6pt] (111,-22.5) -- (114,-22.5);

\node[stepno] at (4.5,-4.5) {1};
\node[head] at (7.5,-4.5) {Approval};

\node[card, line width=1.1pt, draw=black!85] (a1) at (10,-10) {\makebox[37mm][l]{deploy.yml}};
\node[card] (a2) at (10,-19.5) {\makebox[37mm][l]{terraform/main.tf}};
\node[card] (a3) at (10,-29) {\makebox[37mm][l]{terraform/modules/iam/main.tf}};
\draw[tree,->] ([xshift=6mm]a1.south west) -- ([xshift=6mm]a2.north west) node[via, pos=0.5, right=1mm] {runs \texttt{terraform apply}};
\draw[tree,->] ([xshift=6mm]a2.south west) -- ([xshift=6mm]a3.north west) node[via, pos=0.5, right=1mm] {loads module \texttt{iam}};
\draw[black!85, line width=1.6pt] (6.2,-10) -- (6.2,-15.6);
\draw[black!45, line width=0.6pt] (4,-10) -- (3,-10) -- (3,-34.6) -- (4,-34.6);
\draw[black!85, line width=1.6pt] (4,-39.2) -- (4,-41.6);
\node[note, anchor=west] at (5.5,-40.4) {single-object record: \texttt{deploy.yml} only};
\draw[black!45, line width=0.6pt] (4.5,-36) -- (3.5,-36) -- (3.5,-37.6) -- (4.5,-37.6);
\node[note, anchor=west] at (5.5,-36.8) {authority-set record: all three files};

\node[stepno] at (61.5,-4.5) {2};
\node[head] at (64.5,-4.5) {Edit after approval};
\node[card] (b1) at (64,-10) {\makebox[37mm][l]{deploy.yml\hfill{\sffamily\color{black!55}unchanged}}};
\node[card] (b2) at (64,-19.5) {\makebox[37mm][l]{terraform/main.tf\hfill{\sffamily\color{black!55}unchanged}}};
\node[card, draw=accent, line width=1.1pt, fill=accent!7] (b3) at (64,-29) {\makebox[37mm][l]{terraform/modules/iam/main.tf}};
\draw[tree,->] ([xshift=6mm]b1.south west) -- ([xshift=6mm]b2.north west);
\draw[tree,->] ([xshift=6mm]b2.south west) -- ([xshift=6mm]b3.north west);
\node[anchor=north west, font=\ttfamily\scriptsize, text=accent, inner sep=0pt] at (64.4,-35.6) {+ Action = ["s3:PutObject"]};
\node[note, text width=41mm] at (64,-39) {\textcolor{accent}{\textbf{Edited:}} the deployed role gains write access to a bucket.};

\node[stepno] at (118.5,-4.5) {3};
\node[head] at (121.5,-4.5) {Check, then run};
\draw[verdict] (118,-9) rectangle (161,-25.5);
\node[font=\sffamily\bfseries\scriptsize, anchor=north west, inner sep=0pt] at (120,-11) {Single-object check};
\node[badge, fill=accent] at (159.3,-10.3) {ACCEPTS};
\node[note, text width=39mm] at (120,-15.3) {\texttt{H(deploy.yml)} equals the recorded digest.};
\node[note, text width=39mm, text=accent, font=\sffamily\bfseries\scriptsize] at (120,-21.6) {The run applies the edit.};
\draw[verdict] (118,-27.5) rectangle (161,-42);
\node[font=\sffamily\bfseries\scriptsize, anchor=north west, inner sep=0pt] at (120,-29.5) {Authority-set check};
\node[badge, fill=black!80] at (159.3,-28.8) {REJECTS};
\node[note, text width=39mm] at (120,-33.8) {The digest over the three files differs from the record.};
\node[note, text width=39mm, font=\sffamily\bfseries\scriptsize, text=black!85] at (120,-39.6) {Nothing runs.};
\end{tikzpicture}
\caption{Delegated-authority omission: the approval record binds the file presented for review, and the run also reads the files that file leads to. An edit to one of them after approval passes a check on the approved file and fails a check on the authority set.}
\label{fig:attack}
\end{figure*}
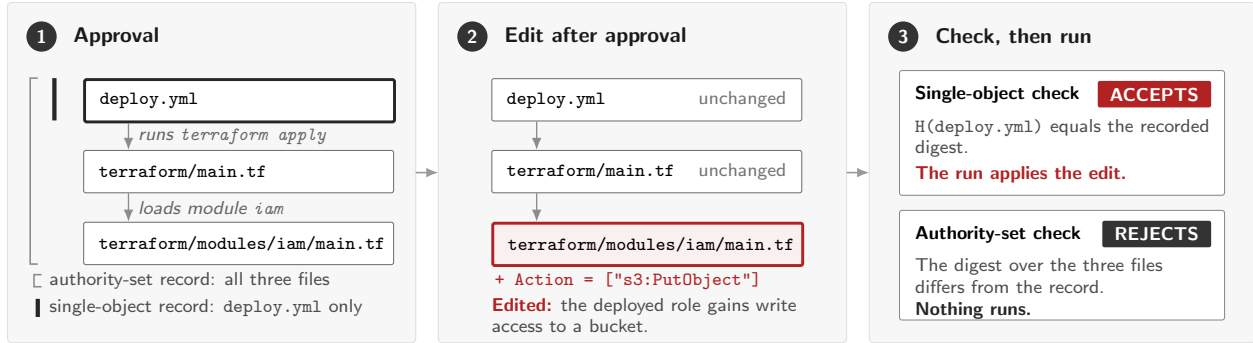

The recorded object is one file or one command, and the run reads more. Consider a workflow, \texttt{deploy.yml}, that runs \texttt{terraform apply}, a command that creates the cloud resources described in a directory of configuration files. One of those files loads a second directory, \nolinkurl{modules/iam}. An approval of \texttt{deploy.yml} is recorded as its path and SHA-256 digest. Before the run starts, \nolinkurl{terraform/modules/iam/main.tf} gains a statement that lets the deployed role write to a storage bucket. The checker hashes \texttt{deploy.yml}, finds the recorded digest, and accepts. The run deploys permissions that the approval never covered (Figure~\ref{fig:attack}). The window between approval and run is open wherever the tree at execution can differ from the tree at approval: a job that checks out a branch name after its reviewer acts, a command that an agent runs later in a session, an allow rule that stands for weeks.

Shipped systems bind approvals to single objects. After a time-of-check-to-time-of-use advisory, one agent runtime refuses interpreter and runtime commands unless the approval binds exactly one concrete local script file~\cite{openclaw2026ghsaxf99}. An editor that approved a tool-server configuration by its entry accepted later changes to the entry's content without a new prompt~\cite{cursor2025cve54136}. Standing approvals in the two agent runtimes we study match the text of a command and bind no file content at all~\cite{anthropic_claude_code_permissions,openai_codex_rules}. Approval tokens for agent runtimes bind the tool and its arguments and leave open the cases in which every recorded field is unchanged and the downstream effect differs~\cite{wang2026approvallaundering}. Continuous-integration security guidance describes the corresponding attack: an attacker who cannot change the pipeline definition edits a file that the definition references~\cite{owasp_cicd_sec4}. Such an attack corrupts nothing the approver saw. It succeeds against an accurate approval of the entrypoint.

One fix is to approve a whole snapshot of the repository, a commit, and run exactly that snapshot. This closes the gap wherever a run can be pinned to a commit. It binds every file in the tree, so any later change, related or not, ends the approval, and it cannot express an approval meant to stand across changes, such as an agent allow rule or a trusted configuration entry. For those approvals the binding has to be scoped to the task. The \emph{authority set} of a task is the set of repository files its run is configured to read, and an approval has \emph{authority-set integrity} when it binds that set.

We measure both halves of the problem. The first is how often the files behind an approved file change while that file stays the same. In 5,251 commits across 20 repositories, the approved file, the \emph{entrypoint}, is unchanged in 296 of the 542 commits that change a file its run reads. Such commits occur in 15 of the 20 repositories as ordinary development. A standing approval bound to the entrypoint's digest would be carried across every one of them, and an allow rule that matches command text binds no file, so no commit ends it. Bound to the authority set, the approval is re-requested on about one commit in ten and stays in force across the 4,709 commits that modify no authority file. The second half is who can name the set. Agent runtimes already rely on the agent to describe the action a user approves~\cite{weng2026consent}, so we give Codex CLI and Claude Code an approval record that can hold a file set and compare what they bind with sets we audited by hand. Their records cover every audited file in the workspace in 59 of 62 runs. Two of the remaining three bind a Terraform deployment and no Terraform file, and nothing in the record distinguishes them from the others.

We make four contributions:
\begin{itemize}[leftmargin=*]
    \item A definition of authority-set integrity, with soundness and exactness conditions for single-object, authority-set, manifest, and commit bindings (Section~\ref{sec:model}).
    \item A corpus of 20 cases from public repositories across six delegation patterns, each with an audited authority set at a pinned revision, and a measurement over their history: 54.6\% of the commits that change an authority file leave the entrypoint unchanged (Sections~\ref{sec:corpus} and~\ref{sec:history}).
    \item A study of two coding-agent runtimes as producers of authority-set records, which locates the difficulty in enumeration rather than hashing: the digest is correct in 61 of 62 records, and the runtimes bind up to five different sets for one task (Section~\ref{sec:agents}).
    \item A receipt checker and three demonstrations in which a post-approval edit to a delegated file passes the single-object check and fails the receipt. One edit adds a bucket-write policy to a role defined in a public repository's Terraform files, and another lowers the required human approvals for high-risk changes in a public repository's policy configuration from 1 to 0, each under an unchanged, approved file (Section~\ref{sec:enforcement}).
\end{itemize}
An approval gate should compute the digest itself and check the set against the tool's own inputs, a declared manifest, or a directory rule, with the agent's set as a proposal (Section~\ref{sec:implications}).

\section{Background}
\label{sec:background}

\paragraph{Repositories and digests.} A repository holds a project's files and their history. A \emph{commit} is a recorded snapshot of every file. The \emph{working tree} is the editable copy that tools and people change between commits. \emph{Checking out} a revision copies its files into the working tree. A \emph{digest} is a short fingerprint of a file's bytes, here SHA-256: any change to the file changes the digest, so comparing digests shows whether a file is the one that was seen earlier. A record that stores a file's digest \emph{binds} that file.

\paragraph{Workflows.} Projects automate building, testing, releasing, and deploying with \emph{workflows}: files in the repository that list the commands an automation service runs. A workflow often does not contain the whole operation. It calls scripts, infrastructure definitions, and configuration stored elsewhere in the repository. Listing~\ref{lst:example} shows the pattern. The workflow says only \texttt{terraform apply}. What that command creates, including the cloud permissions it grants, is written in the Terraform files the command loads. We call the file presented for approval the \emph{entrypoint} and the files it leads to \emph{delegated} files.

\begin{lstlisting}[caption={A simplified deployment workflow and one of the files it delegates to. The permissions the deployment grants are set in the second file.},label={lst:example}]
# .github/workflows/deploy.yml        (the file presented for approval)
jobs:
  deploy:
    steps:
      - uses: actions/checkout@v4
      - run: terraform apply -auto-approve
        working-directory: terraform

# terraform/main.tf                   (loaded by the command above)
module "iam" {
  source = "./modules/iam"            # which loads terraform/modules/iam/main.tf
}
\end{lstlisting}

\paragraph{Tools in the examples.} Four kinds of delegated file appear in the examples. \emph{Terraform} describes cloud infrastructure in text files, grouped into directories called modules, and a deployment creates whatever those files describe. \emph{Helm} installs software on a cluster from a \emph{chart}, a directory of templates and values files, and each installation is a release. A \emph{policy module} is a small program that decides whether a change is allowed, and its thresholds can be kept in a separate configuration file. Rego is a language for writing such modules. A \emph{local action} is a workflow step defined in the same repository and called by path.

\paragraph{Approvals.} Because such operations change production systems, they are gated. A deployment waits until a designated reviewer approves it. A coding agent, an AI assistant that edits files and runs commands in a repository on a user's behalf, pauses and asks before running a command. The program that runs the agent, shows the request, and executes the command is the agent \emph{runtime}. A user can also grant a standing approval, an \emph{allow rule}, that lets a matching command run again without asking. The approval is stored as a record of what was approved, and a check before execution compares the record with what is about to run. Whether that check means what the approver intended depends on what the record contains.

\section{Setting and Threat Model}
\label{sec:threat}

\paragraph{Roles.} An \emph{approval producer} emits an \emph{approval record} for a task in a repository. The producer is a human operator, a policy bot, or a coding agent that describes a pending action for its user. A \emph{checker} validates the record against the repository state before execution, and an \emph{executor} runs the workflow, script, policy evaluation, or command. The checker enforces what the record binds and nothing else.

\paragraph{Approvals that are not pinned to a commit.} Agent runtimes approve an operation and dispatch it later, and the approved script, the workflow state, or the dispatched action can change in between~\cite{openclaw2026ghsaxf99,kumar2026loopjacking,wang2026approvallaundering}. Configuration approvals trust an entry once and accept later changes to its content without a new prompt~\cite{cursor2025cve54136}. Some review-gated workflows check out a branch name after the reviewer acts. They run whatever that name points to at checkout time~\cite{lobacevski2021pwnrequests,khan_actionstoctou}. In each design the tree at execution time can differ from the tree at approval time.

\paragraph{Adversary.} The adversary changes a repository file after the approval record is produced and before the run starts. The file is one the run reads and the record does not bind: a module owned by another team, a values file regenerated by automation, a script the agent edits later in the session. The adversary needs only the write access that an ordinary contributor to that file already has. The adversary cannot find SHA-256 collisions, alter the stored record, or compromise the checker, the executor, or state outside the repository. The adversary succeeds when the checker accepts and the run reads the changed file.

\paragraph{Vulnerability class.} A \emph{delegated-authority omission} is an approval record that binds the file presented for review while the run reads further repository files outside that binding. The record is authentic and the approved file is unmodified. The attack needs no injected command and no forged approval: it changes a file the record never named. It is the approval-record form of indirect poisoned pipeline execution~\cite{owasp_cicd_sec4}, in which the attacker edits a file the pipeline definition references.

\paragraph{Security property.} The run reads only repository content that the approval bound. To hold at execution time, the executor must read the content the checker verified, for example by running from the verified snapshot. The property covers the files in the repository tree at the time of the check. Runner images, package registries, third-party actions, cloud state, and secrets are governed by their own controls, such as pinning actions to commit SHAs~\cite{github_secure_use}, a lockfile of a workflow's resolved action graph~\cite{github_actions_lockfile}, and artifact attestations~\cite{github_attestations}.

\paragraph{Commit binding.} Executing exactly the approved commit satisfies the property for every task, and it is the recommended fix for review-gated workflows~\cite{codeql_untrusted_checkout_toctou}. A hosted workflow run carries the commit that triggered it, and a job that checks out that commit runs under a commit binding. An agent's working tree can be snapshotted the same way, at the same cost: any later edit ends the approval. A standing approval has no snapshot to pin to.

\section{Authority-Set Integrity}
\label{sec:model}

Let $R$ be a repository state and $t$ a task: a command or workflow invocation with its parameters. The \emph{authority set} $A(R,t)$ is the set of paths in $R$ that the run of $t$ is configured to read. It contains the \emph{entrypoint} $e$ presented for approval and every path reached from it through local references: an invoked script, a local action, a module source, a chart's files, a policy's imported modules and data. The members other than $e$ are \emph{delegated files}, and we call any member an \emph{authority file}. The definition is static. It follows what the configuration names, and a particular execution may read fewer files.

A binding $B$ is the commitment stored in an approval record. A checker $V(B,R')$ accepts or rejects it against the state $R'$ found at execution time.

\paragraph{Single-object binding.} $B=(p,\sha(R[p]))$ for one path $p$. The checker accepts when $\sha(R'[p])$ equals the recorded digest.

\paragraph{Authority-set binding.} $B=(S,d)$ for a set $S$ of entries and a canonical digest $d$ over them (Listing~\ref{lst:digest}). The checker recomputes $d$ over $R'$ and accepts on equality. An entry is a file or a directory. A file entry binds the file's content. A directory entry binds the sorted names and digests of the directory's members, for tools that load every file in a directory, such as a Terraform module.

\begin{lstlisting}[caption={Canonical authority-set digest over file entries.},label={lst:digest}]
records = []
for path in sorted(authority_paths):
    records.append(path + "\0" + sha256(read_bytes(path)))
authority_set_sha256 = sha256("\n".join(records))
\end{lstlisting}

\paragraph{Manifest binding.} $B=(m,\sha(R[m]))$ for a manifest $m$ that lists entries and digests. The checker verifies the digest of $m$ and then each listed digest against $R'$. Lockfiles have this form.

\paragraph{Commit binding.} $B$ is a commit identifier, and the executor runs that commit's tree.

\paragraph{Soundness and exactness.} A binding is \emph{sound} for $(R,t)$ when every member of $A(R,t)$ is bound, directly or through a verified manifest, and every directory that a tool loads in full is bound as a directory entry. If $B$ is sound and $V$ accepts in $R'$, every member of the authority set has the content that was approved: a modified or deleted file changes a file digest, and a file added to a loaded directory changes a directory digest. A binding is \emph{exact} when it is sound and binds nothing outside $A(R,t)$, so that it is invalidated only by a change to something the run is configured to read. An approval has \emph{authority-set integrity} when its binding is sound.

A single-object binding on $e$ is sound only when $A(R,t)=\{e\}$. A commit binding is sound for every task and exact only for tasks whose authority set is the whole tree. An authority-set or manifest binding is as good as its set: sound when $S \supseteq A(R,t)$ and exact when $S = A(R,t)$. The digest is mechanical. Naming $A(R,t)$ is the hard part, for auditors (Section~\ref{sec:corpus}) and for agents (Section~\ref{sec:agents}).

\paragraph{Receipts.} A \emph{receipt} is an authority-set binding whose digest the checker computes. The producer may propose the set, and the checker hashes the files itself.

\section{Corpus}
\label{sec:corpus}

\paragraph{Collection.} We queried GitHub code search with 31 queries written for specific delegation mechanisms, for example \texttt{id-token: write} under \nolinkurl{.github/workflows}, \texttt{helm values-prod}, \texttt{rego policy override}, and \nolinkurl{uses: ./.github/actions}. The queries returned 141 candidate files in 139 repositories. A candidate is a file that could be presented for approval, the entrypoint, at a fixed revision. We built a candidate authority set for 30 of them, by a script that follows local references from the entrypoint and by hand.

\paragraph{Audit.} We audited each of the 30 candidate sets edge by edge against file content, where an edge is a reference from one file to another. A file stays in a set when a reference from the entrypoint or from another member reaches it and it determines what the run deploys, publishes, or decides. The audited set is thus the set of files that decide the outcome of the run. Neighbors in the tree, sibling workflows, documentation, tests, and examples are removed, and referenced files that the candidate set missed are added. For the Helm case, for example, the audit kept the production workflow, the chart metadata, values, and rendered templates, the deploy script, and a cluster-role manifest, and it removed documentation, notes, test templates, examples, development values, and a notification action. Naming the set was hard for the auditors too. The audit left 8 of the 30 sets unchanged, changed 12, and excluded 10: one with no delegated edge, two that depend on content outside the repository, and seven whose sets it could not complete. Two other candidates, outside the 30 and outside the corpus, reduced to the entrypoint alone under the same audit, and one of them serves as a single-file control in Section~\ref{sec:agents}.

\paragraph{Result.} The corpus holds 20 cases in six delegation patterns (Table~\ref{tab:patterns}). Each case records the repository, the full revision hash, the entrypoint, every authority file with its source URL and SHA-256 digest, and the canonical set digest over file entries. The audited sets list the files present at the pinned revision, and the evaluation scores records at that file level. A deployed receipt for a tool that loads whole directories, such as a Terraform module, adds a directory entry for each, so that a file added to the directory also invalidates it. Authority sets range from 2 to 44 files, with a median of 5 and a total of 215. Every case delegates, and the cases were chosen to cover distinct delegation mechanisms. Sixteen entrypoints are files under \nolinkurl{.github} (workflows, a workflow generator, and an action), three are policy modules, and one is a Kubernetes manifest. The delegation they exhibit is a property of the repository and holds for any approval of the same entrypoint, whether a reviewer, a bot, or an agent runtime issues it.

\begin{table*}[t]
\centering
\caption{Every case in the corpus delegates from its entrypoint to further repository files, through one of six mechanisms. Files: authority files summed over the pattern's cases.}
\label{tab:patterns}
\small
\renewcommand{\arraystretch}{1.2}
\begin{tabular}{L{0.15\linewidth}L{0.35\linewidth}L{0.28\linewidth}rr}
\toprule
Pattern & Delegated files & Example (entrypoint $\rightarrow$ delegated) & Cases & Files \\
\midrule
Permission scope & Terraform modules and Kubernetes manifests that set cloud roles, workflow permissions, or cluster access rules & \nolinkurl{loxilb-io/loxilb}: \texttt{eks.yaml} $\rightarrow$ \nolinkurl{cicd/eks/main.tf}, manifests & 5 & 85 \tabularnewline
Environment routing & Chart or infrastructure definitions, values, templates, and scripts that set what a release deploys & \nolinkurl{nc2U/ibs}: \texttt{helm\_prod.yml} $\rightarrow$ chart, values, deploy script & 3 & 78 \tabularnewline
Package and provenance & Build definitions and release manifests that set what is packaged and published & \nolinkurl{perkeep/perkeep}: \texttt{docker.yml} $\rightarrow$ \texttt{Dockerfile} & 5 & 18 \tabularnewline
Cross-file policy & Configuration, data files, and other modules that a policy decision reads & \nolinkurl{Agnostic-Security/yashigani}: \texttt{yashigani.rego} $\rightarrow$ two Rego modules, access-rule data & 3 & 11 \tabularnewline
Generated workflows & A generator and the workflow files it emits, which CI runs & \nolinkurl{rauc/meta-rauc-community}: \texttt{generate.py} $\rightarrow$ five workflow files & 2 & 13 \tabularnewline
Local actions & Nested local actions, lock files, and configuration that a workflow or action reads & \nolinkurl{bitcoin/bitcoin}: \nolinkurl{cache/save/action.yml} $\rightarrow$ \nolinkurl{internal/action.yml} & 2 & 10 \tabularnewline
\midrule
Total & & & 20 & 215 \tabularnewline
\bottomrule
\end{tabular}
\end{table*}

\section{Evaluation}

\subsection{A digest of the entrypoint passes 296 of the 542 commits that change what the run reads}
\label{sec:history}

A single-object binding misses a change when a delegated file changes and the entrypoint does not. We count those changes in the history of the 20 corpus repositories.

\paragraph{Method.} For each case we take up to 500 commits ending at the pinned revision, excluding merge commits, and keep those in which the entrypoint exists, which leaves 4 to 500 commits per repository and 5,251 in total. For each commit we record whether it modifies the entrypoint and whether it modifies any authority file, with paths matched as named at the pinned revision. A \emph{delegated-only} commit modifies an authority file and leaves the entrypoint unchanged.

\paragraph{The entrypoint is unchanged in 296 of the 542 commits that modify an authority file.} Of 5,251 commits, 542 modify an authority file and 246 modify the entrypoint (Table~\ref{tab:history}). The other 296, 54.6\% of the authority-changing commits, are delegated-only, and an entrypoint digest is the same before and after every one of them. They occur in 15 of the 20 repositories in the ordinary course of development, so the state a single-object check cannot see arises with no adversary. With any one repository removed, the share stays between 43.6\% and 61.9\%. Commits are nested in repositories, and resampling the 20 repositories gives a 95\% interval of 35.6\% to 70.6\% for the pooled share.

\begin{figure}[t]
\centering
\begin{tikzpicture}
\begin{axis}[xbar stacked, width=0.78\linewidth, height=5.4cm, bar width=8pt,
  xmin=0, xmax=262, ymin=0.4, ymax=6.6, y dir=reverse, xlabel={Commits that change an authority file}, xlabel style={font=\footnotesize},
  xtick={0,50,100,150,200},
  ytick={1,2,3,4,5,6}, yticklabels={Environment routing,Package and provenance,Permission scope,Cross-file policy,Local actions,Generated workflows},
  yticklabel style={font=\footnotesize}, xticklabel style={font=\footnotesize}, ytick style={draw=none},
  legend style={at={(0.5,1.04)}, anchor=south, font=\footnotesize, draw=none, legend columns=2, /tikz/every even column/.append style={column sep=8pt}}, legend cell align=left,
  axis lines*=left]
\addplot+[fill=black!75, draw=black] coordinates {(145,1) (46,2) (56,3) (7,4) (41,5) (1,6)};
\addplot+[fill=white, draw=black] coordinates {(39,1) (35,2) (46,3) (13,4) (90,5) (23,6)};
\legend{Delegated-only (entrypoint unchanged), Entrypoint changed}
\node[font=\footnotesize, anchor=west] at (axis cs:186,1) {145 of 184 (78.8\%)};
\node[font=\footnotesize, anchor=west] at (axis cs:83,2) {46 of 81 (56.8\%)};
\node[font=\footnotesize, anchor=west] at (axis cs:104,3) {56 of 102 (54.9\%)};
\node[font=\footnotesize, anchor=west] at (axis cs:22,4) {7 of 20 (35.0\%)};
\node[font=\footnotesize, anchor=west] at (axis cs:133,5) {41 of 131 (31.3\%)};
\node[font=\footnotesize, anchor=west] at (axis cs:26,6) {1 of 24 (4.2\%)};
\end{axis}
\end{tikzpicture}
\caption{An entrypoint digest misses more than half of the authority-changing commits in three of the six patterns. The dark segment of each bar is the part it misses, and the label gives that segment as a count and a share of the bar.}
\label{fig:history}
\end{figure}
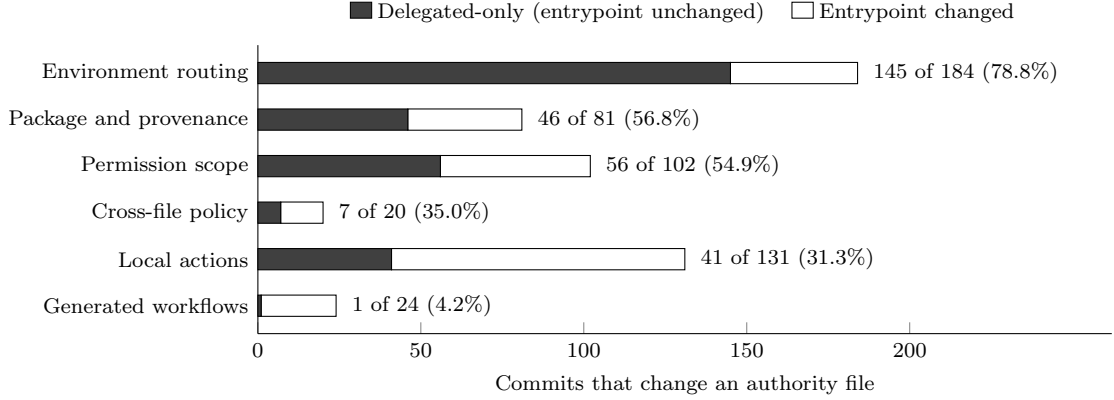

\paragraph{The share runs from 1 of 24 for generated workflows to 145 of 184 for environment routing.} The share differs by pattern (Figure~\ref{fig:history}). A generator and the workflows it emits change together, while the charts, values, and infrastructure definitions behind a deployment workflow change far more often than the workflow. The counts are concentrated. Three repositories contribute 203 of the 296 delegated-only commits, led by one environment-routing repository, which deploys with Terraform, with 130. Without that repository the pooled share is 166 of 381 (43.6\%), and the other two environment-routing repositories still show 15 of 23. Across the 13 repositories with at least ten authority-changing commits, the share runs from 0\% to 92\%, and 6 of them are at or above 50\%.

\paragraph{A standing approval bound to the authority set is re-requested on about one commit in ten.} An approval that is meant to remain valid while the repository changes, such as an allow rule for a deployment command, is re-requested under an authority-set binding on the 542 commits that modify an authority file, about one in ten, and stays in force across the other 4,709. Under a single-object binding it is re-requested on 246 commits and silently carried across the 296 that change what the run reads. An allow rule that matches command text binds no file, and no commit to these files ends it. A commit binding ends at the next commit and does not express a standing approval. The 296 extra requests of the authority-set binding are exactly the commits the single-object binding lets through.

\begin{table}[t]
\centering
\caption{The entrypoint is unchanged in 296 of the 542 commits that modify an authority file. Counts are commits that modify the entrypoint or an authority file, over commits in which the entrypoint exists. The entrypoint is itself an authority file, so the authority-file count includes commits that modify it. Delegated-only commits modify an authority file and leave the entrypoint unchanged.}
\label{tab:history}
\small
\setlength{\tabcolsep}{4pt}
\begin{tabular}{L{0.24\linewidth}rrrr}
\toprule
Pattern & Commits & Entrypoint & Authority file & Delegated-only \\
\midrule
Permission scope & 1,571 & 46 & 102 & 56 \tabularnewline
Environment routing & 1,054 & 39 & 184 & 145 \tabularnewline
Package and provenance & 727 & 35 & 81 & 46 \tabularnewline
Cross-file policy & 1,034 & 13 & 20 & 7 \tabularnewline
Generated workflows & 588 & 23 & 24 & 1 \tabularnewline
Local actions & 277 & 90 & 131 & 41 \tabularnewline
\midrule
Total & 5,251 & 246 & 542 & 296 \tabularnewline
\bottomrule
\end{tabular}
\end{table}

\subsection{Two agent runtimes compute the digest correctly and bind different sets}
\label{sec:agents}

An authority-set binding needs someone to supply the set. In an agent runtime the natural candidate is the agent, which has read the repository and already writes the description the user approves. Two runtimes in that role compute the digest correctly in 61 of 62 records and bind five different file sets for one Terraform deployment.

\paragraph{Cases.} Three cases come from the corpus and delegate: a Terraform deployment (\nolinkurl{devsecblueprint/devsecblueprint}), a Helm deployment (\nolinkurl{nc2U/ibs}), and a policy decision (\nolinkurl{itdojp/ae-framework}). The fourth case is the single-file control, a local action (\nolinkurl{pyca/cryptography}) that the audit reduced to the action file alone. Each workspace is a copy of part of the repository at the pinned revision, holding 50, 74, 7, and 80 repository files and two task-description files. The Terraform and Helm workspaces contain 43 of the 44 audited authority files, because one file in each set, a task runner and a cluster-role manifest, lies outside the copied directories. Scoring uses the audited files present in the workspace: 43, 43, 2, and 1. In the Terraform workspace the workflow installs Terraform and runs the deployment task, the Terraform directory is present, and the task runner that connects the two is the file outside it.

\paragraph{Record schemas.} The task names the entrypoint as the file under review and asks for an approval summary and an approval record with two fields, a name and a SHA-256 digest. Under the \emph{single-object} schema the record holds one path and that file's digest. The \emph{abstention} schema is the same record with one addition: the agent writes \texttt{SCHEMA\_INSUFFICIENT} if the run depends on more than one file. Under the \emph{file-set} schema the prompt says that if the authority for the run is a file set, the record holds the name \texttt{FILE\_SET} and the canonical digest of Listing~\ref{lst:digest}, whose construction it states. Which files belong is left to the agent, as it is whenever an agent describes an action for approval. A fourth, \emph{prose} schema keeps the single-object record and asks the summary to describe the files the run depends on. A file-set record holds a digest and no list of files, so we identify the set it binds by recomputing the digest over candidate sets of workspace files (the audited files, the entrypoint with each directory, every repository file, and those with the task files) until one matches. Every record matches one candidate, 61 under the canonical construction and one under \nolinkurl{./}-prefixed paths. A wrong hash computation matches no candidate, so a match shows both the set and a correct digest of it. A record \emph{covers} the audited set when its digest is the canonical digest of a set that contains every audited file in the workspace, and is \emph{exact} when that set contains no other file.

\paragraph{Runtimes.} We ran OpenAI Codex CLI 0.142.4 with \texttt{gpt-5.5}, reasoning effort set to none, and Claude Code 2.1.196 with \texttt{claude-sonnet-4-6} at its default settings, both with command approval disabled, in June 2026. Each runtime ran each case under each schema 8 times, for 256 runs, and every repetition of a case and schema started from byte-identical workspace files and the same prompt. The audited sets were in no file an agent could reach. The directory above each workspace held an earlier four-file reference set for the case, which differs from the audited set in all four cases, so reading it cannot produce an exact record. In two Codex runs, both on the policy case under the file-set schema, the agent read that directory, and both records bind the four-file set. Both runs are excluded, which leaves 254 runs and 62 under the file-set schema. No other file-set record from either runtime binds that set.

\paragraph{A record with room for one file binds the entrypoint.} Under the single-object schema and under the prose schema, all 128 records hold the entrypoint and its correct digest. For the control that record is exact. For the delegating cases it binds 1 of 43, 1 of 43, and 1 of 2 authority files, in the 48 prose-schema runs as in the others.

\paragraph{Abstention does not track delegation.} On the single-file control the runtimes wrote \texttt{SCHEMA\_INSUFFICIENT} in 15 of 16 runs, against 44 of 48 on the delegating cases (Table~\ref{tab:agents}). The audited authority of the control is one file, and the single-object record suffices for it. Claude Code bound the policy entrypoint without abstaining in 4 of its 8 runs on that case, the one with the smallest delegated set. Under the file-set schema the same runtimes bound the action file alone in 13 of the 16 control runs. The flag fires on a single-file task as often as on a delegating one, so it cannot route approvals between a single-object check and a set check.

\paragraph{File-set records cover the audited files in 59 of 62 runs.} Thirty-five of the 59 bind a strict superset, and 24 are exact: Codex on the Terraform case (8 of 8), the policy case (3 of 6), and the control (6 of 8), and Claude Code on the control (7 of 8). The supersets are mostly directories. Both runtimes bound the workflow plus every file under \nolinkurl{deploy/helm} in all 16 Helm runs, 62 files against the audited 43. The 19 extra files are ignore files, notes, test templates, documentation, an example manifest, development values, and other scripts in the chart directory. Helm loads the chart directory itself, under the control of its ignore files, so the directory is the set that the tool opens. Against that set those 16 records are exact and the exact count is 40 of 62. On the policy case 11 records bound all seven files of the policy directory, and five of those added the two task-description files. On the Terraform case two Claude Code records bound everything under \texttt{terraform}, and three bound the entire workspace. On the control three records added other files from the repository.

\paragraph{Two records bind a Terraform deployment and no Terraform file.} Three records do not cover the audited set. Two Claude Code records for the Terraform case bind five workflow files and none of the 42 Terraform files. Both are well formed, and both carry a correct digest over the five files. A checker that enforces either one accepts any change to the Terraform configuration. The summaries written with both records say that the workflow uses Terraform, and from the same workspace and prompt Codex bound all 42 Terraform files in 8 of 8 runs. The third record's digest is computed over every repository file with \nolinkurl{./}-prefixed paths, so it does not match the canonical construction and a checker that recomputes it rejects the record.

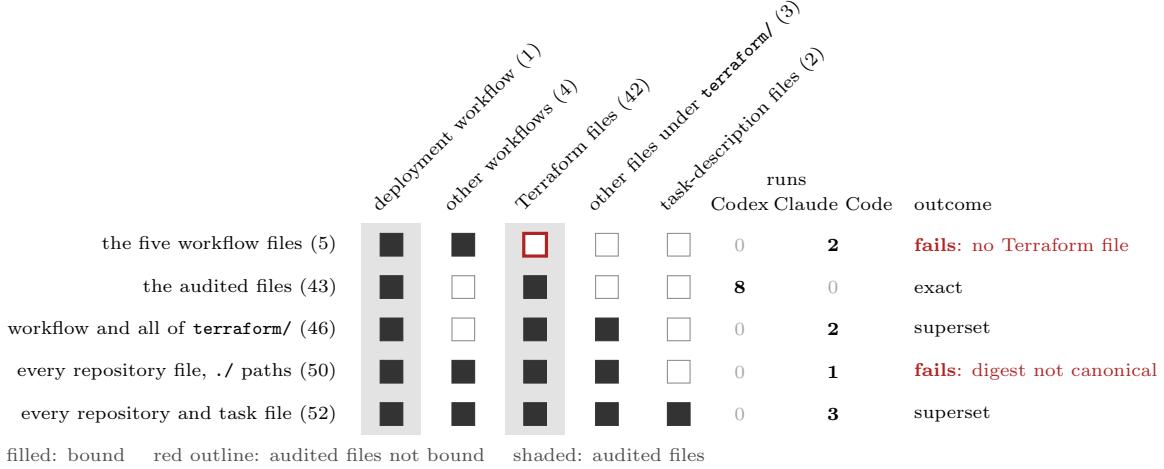
\begin{figure}[t]
\centering
\begin{tikzpicture}[font=\scriptsize, x=1cm, y=1cm,
  cellon/.style={fill=black!80, draw=black!80, minimum size=3mm, inner sep=0pt},
  celloff/.style={fill=white, draw=black!45, minimum size=3mm, inner sep=0pt},
  cellmiss/.style={fill=white, draw=accent, very thick, minimum size=3mm, inner sep=0pt},
  rowlab/.style={anchor=east, align=right, font=\scriptsize},
  collab/.style={anchor=south west, rotate=45, font=\scriptsize, inner sep=1pt},
  runs/.style={anchor=center, font=\scriptsize},
  outlab/.style={anchor=west, font=\scriptsize}]
\def\dx{0.95}
\def\dy{0.56}
\fill[gray!22] ($(1*\dx-0.5*\dx+0.08, 0.5*\dy)$) rectangle ($(1*\dx+0.5*\dx-0.08, -4.5*\dy)$);
\fill[gray!22] ($(3*\dx-0.5*\dx+0.08, 0.5*\dy)$) rectangle ($(3*\dx+0.5*\dx-0.08, -4.5*\dy)$);
\node[collab] at (1*\dx-0.12, 0.5*\dy+0.05) {deployment workflow (1)};
\node[collab] at (2*\dx-0.12, 0.5*\dy+0.05) {other workflows (4)};
\node[collab] at (3*\dx-0.12, 0.5*\dy+0.05) {Terraform files (42)};
\node[collab] at (4*\dx-0.12, 0.5*\dy+0.05) {other files under \texttt{terraform/} (3)};
\node[collab] at (5*\dx-0.12, 0.5*\dy+0.05) {task-description files (2)};
\node[anchor=south, font=\scriptsize] at (6.5*\dx, 0.5*\dy+0.38) {runs};
\node[anchor=south, font=\scriptsize] at (5.85*\dx, 0.5*\dy+0.05) {Codex};
\node[anchor=south, font=\scriptsize] at (7.15*\dx, 0.5*\dy+0.05) {Claude Code};
\node[anchor=south west, font=\scriptsize] at (8.15*\dx, 0.5*\dy+0.05) {outcome};
\foreach \r/\lab/\cA/\cB/\cC/\cD/\cE/\codex/\claude/\outc/\col in {
  0/{the five workflow files (5)}/1/1/2/0/0/0/2/{\textbf{fails}: no Terraform file}/accent,
  1/{the audited files (43)}/1/0/1/0/0/8/0/{exact}/black,
  2/{workflow and all of \texttt{terraform/} (46)}/1/0/1/1/0/0/2/{superset}/black,
  3/{every repository file, \texttt{./} paths (50)}/1/1/1/1/0/0/1/{\textbf{fails}: digest not canonical}/accent,
  4/{every repository and task file (52)}/1/1/1/1/1/0/3/{superset}/black}{
  \node[rowlab] at (0.5*\dx-0.12, -\r*\dy) {\lab};
  \foreach \c/\v in {1/\cA,2/\cB,3/\cC,4/\cD,5/\cE}{
    \ifnum\v=1 \node[cellon] at (\c*\dx, -\r*\dy) {}; \fi
    \ifnum\v=0 \node[celloff] at (\c*\dx, -\r*\dy) {}; \fi
    \ifnum\v=2 \node[cellmiss] at (\c*\dx, -\r*\dy) {}; \fi
  }
  \node[runs] at (5.85*\dx, -\r*\dy) {\ifnum\codex=0 \textcolor{black!35}{0}\else\textbf{\codex}\fi};
  \node[runs] at (7.15*\dx, -\r*\dy) {\ifnum\claude=0 \textcolor{black!35}{0}\else\textbf{\claude}\fi};
  \node[outlab, text=\col] at (8.15*\dx, -\r*\dy) {\outc};
}
\node[anchor=north east, font=\scriptsize, text=black!70, align=right] at (5.5*\dx, -4.5*\dy-0.05) {filled: bound \quad red outline: audited files not bound \quad shaded: audited files};
\end{tikzpicture}
\caption{Sixteen runs on the Terraform case, from the same workspace and prompt, bind five different file sets. Codex binds the audited 43 files in all eight runs. Claude Code binds four different sets in eight runs, and two of those runs bind none of the 42 Terraform files. The workspace holds 50 repository files in four groups and two task-description files.}
\label{fig:sets}
\end{figure}

\paragraph{One task draws five different sets.} Sixty-one of the 62 digests are the canonical digest of the set the record binds, so hashing is not where the records differ. The 16 Terraform runs bind five different sets (Figure~\ref{fig:sets}), four of them from one runtime in eight runs. The 14 policy runs bind three sets and the 16 control runs three. Helm is the one case on which every run agrees. The count of distinct sets needs no reference: at most one of five different sets can be exact. A record has the same form whether its set is exact, over-inclusive, or missing every Terraform file, and a checker that accepts the producer's digest enforces all three alike. A record that also listed its files would let the checker recompute the digest and would still not show whether the list is complete.

\begin{table*}[t]
\centering
\caption{The runtimes abstain on the single-file control as often as on the delegating cases, cover the audited files in 59 of 62 file-set runs, and match them exactly in 24. Records from two coding-agent runtimes, 8 runs per cell (6 for Codex on the policy case under the file-set schema). Abstained: wrote \texttt{SCHEMA\_INSUFFICIENT} under the abstention schema. Covers and Exact: file-set records that bind every audited file in the workspace, and that bind exactly those files. Files: audited files present in the workspace. Sets: distinct file sets over both runtimes.}
\label{tab:agents}
\footnotesize
\setlength{\tabcolsep}{2.5pt}
\begin{tabular}{L{0.19\linewidth}rccccccr}
\toprule
 & & \multicolumn{2}{c}{Abstained} & \multicolumn{2}{c}{Covers} & \multicolumn{2}{c}{Exact} & \\
Case & Files & Codex & Claude & Codex & Claude & Codex & Claude & Sets \\
\midrule
Terraform deployment & 43 & 8/8 & 8/8 & 8/8 & 5/8 & 8/8 & 0/8 & 5 \tabularnewline
Helm deployment & 43 & 8/8 & 8/8 & 8/8 & 8/8 & 0/8 & 0/8 & 1 \tabularnewline
Policy decision & 2 & 8/8 & 4/8 & 6/6 & 8/8 & 3/6 & 0/8 & 3 \tabularnewline
Local action (control) & 1 & 8/8 & 7/8 & 8/8 & 8/8 & 6/8 & 7/8 & 3 \tabularnewline
\midrule
Total & & 32/32 & 27/32 & 30/30 & 29/32 & 17/30 & 7/32 & \tabularnewline
\bottomrule
\end{tabular}
\end{table*}

\subsection{A single-object check accepts edits that widen a role and remove a review requirement in two public repositories' files}
\label{sec:enforcement}

A receipt checker stores a receipt at approval and recomputes it over the working tree before execution. A gate built on it runs the approved command only when the check accepts. Three demonstrations carry out the attack of Figure~\ref{fig:attack}. The first executes a command behind the gate. The other two apply an edit to the files of a corpus case and run both checks over the audited set, on local copies.

\paragraph{The gate runs the edited deployment under a single-object binding and refuses it under a receipt.} A local gate approves \texttt{deploy.yml}, which runs \nolinkurl{scripts/deploy.py}, which reads \texttt{policy.json}. The gate stores the approval, checks it against the working tree, and on acceptance executes the command named in \texttt{deploy.yml}. After approval we edit \texttt{policy.json}. Listing~\ref{lst:gate} shows the outcome. With a single-object binding on \texttt{deploy.yml} the gate accepts, and the command it runs now reports \texttt{admin} among the permissions it deploys. With a receipt over the three files the gate rejects and the command does not run.

\noindent\begin{minipage}{\linewidth}
\begin{lstlisting}[caption={Gate behavior before and after a post-approval edit to \texttt{policy.json}. The digest of the approved file is the same in both columns.},label={lst:gate}]
                         before edit              after edit
H(deploy.yml)            30c709dd...              30c709dd...   (unchanged)
single-object check      accept                   accept
  command output         DEPLOY_PERMISSIONS=read  DEPLOY_PERMISSIONS=read,admin
authority-set digest     843072bb...              e250cbcf...
authority-set check      accept                   reject
  command output         DEPLOY_PERMISSIONS=read  (not executed)
\end{lstlisting}
\end{minipage}

\paragraph{A post-approval edit adds a bucket-write policy to a deployed role under an approved workflow.} In the corpus case \nolinkurl{devsecblueprint/devsecblueprint} the approved entrypoint is the deployment workflow, which invokes a task runner that applies the Terraform configuration. After approval we append the block in Listing~\ref{lst:tamper} to the module that configuration loads to define access roles. The role previously held read access to the content bucket, and the block adds write access. The workflow file is byte-identical and the single-object check accepts. The receipt over the 44 audited files rejects.

\noindent\begin{minipage}{\linewidth}
\begin{lstlisting}[caption={Post-approval edit to \texttt{terraform/modules/iam/main.tf}. The approved workflow is not touched.},label={lst:tamper}]
+ resource "aws_iam_role_policy" "extra_s3_write" {
+   name   = "${var.role_name}-extra-s3-write"
+   role   = aws_iam_role.lambda_execution.id
+   policy = jsonencode({
+     Version   = "2012-10-17"
+     Statement = [{
+       Effect   = "Allow"
+       Action   = ["s3:PutObject"]
+       Resource = ["${var.content_registry_bucket_arn}/*"]
+     }]
+   })
+ }
\end{lstlisting}
\end{minipage}

\paragraph{A one-line edit lowers the required human approvals for high-risk changes from 1 to 0 under an approved policy module.} In the corpus case \nolinkurl{itdojp/ae-framework} the approved entrypoint is the policy module \nolinkurl{policy/risk-policy.rego}. The thresholds for its decisions are set in \nolinkurl{policy/risk-policy.yml}, which the repository's policy loader reads. After approval we change one line of that file, from \texttt{min\_human\_approvals: 1} to \texttt{min\_human\_approvals: 0} under \texttt{high\_risk}. The policy module is byte-identical, the single-object check accepts, and the receipt over both files rejects. The repository's own loader returns a required approval count of 1 for the original file and 0 for the edited one.

\section{Security Implications}
\label{sec:implications}

\paragraph{The set must come from a source the checker can verify.} A set named by the approval producer is an assertion, and the agent is a producer. The gate should compute the receipt itself, which removes the one malformed digest we observed, and take the set from a source it can check. The tool can report its inputs: a Terraform plan file embeds the configuration it was computed from, and applying the saved plan performs the planned operations~\cite{hashicorp_terraform_apply,hashicorp_terraform_automation}. The repository can declare the set in a manifest that is itself bound and reviewed. The gate can over-approximate by directory, which is what both runtimes did on every Helm run. A directory rule is sound whenever the delegated files sit under the named directories, and it costs re-approvals for unrelated edits there. In two of the three delegating cases one audited file, a task runner or a cluster-role manifest, sits outside the directories that hold the rest of the set, so a directory rule has to name it separately. The agent's set remains useful as a proposal to test against one of these sources.

\paragraph{A checker that accepts one path and one digest is correct only for single-file tasks.} For any other task it should require an authority-set receipt or a manifest whose listed digests it verifies, and it should run the task from the content it verified. The agent's own flag does not tell the checker which kind of task it faces: the runtimes flagged the single-file control in 15 of 16 runs and bound that file alone in 13 of 16 under another schema.

\paragraph{An entrypoint that delegates is an incomplete approval target, for any control keyed to it.} The history measurement is about where changes land, so it applies to a path-scoped review rule or a change alert as much as to a digest: a control keyed to the entrypoint alone observes 246 of the 542 commits that modify an authority file. A run that can be pinned to a commit should be. An approval that stands across changes needs a task-scoped set, and the price varies with what is delegated, from 20 of 1,034 commits for the policy cases to 131 of 277 for local actions (Table~\ref{tab:history}).

\paragraph{An evaluation of agent-written approval records has to score the set.} A well-formed record with a correct digest says nothing about soundness, so the score is set membership against a reference, over repeated runs: one run per cell would have reported any one of four sets for one runtime on the Terraform case. Coverage is the safety measure and holds at 59 of 62 whichever Helm set is the reference, while exactness moves from 24 to 40. Without the single-file control, 44 abstentions in 48 delegating runs would read as detection.

\section{Scope of the Measurements}

Authority sets follow explicit references, which an auditor can check by reading files. For a tool that takes its inputs in bulk, such as a Docker build that copies its context, the binding that covers the run is a directory entry for the context or the commit.

The audited sets are fixed from file content, edge by edge. Two results hold whichever reference is used for the Helm chart: records cover the reference in 59 of 62 runs, and the number of distinct sets per case is the same. Exactness follows the reference, 24 of 62 against the audited sets and 40 against the chart directory.

The history measurement counts the event a digest check observes, a change to the bytes of a file. Paths are matched by their names at the pinned revision, which applies one authority set per repository across its whole history.

The agent study reports each runtime separately, as we ran it, on four cases with 8 runs per cell. The result a checker depends on is variation within a runtime: Claude Code bound four different sets in eight runs on one workspace and prompt.

\section{Related Work}

\paragraph{Approval binding for agents.} Weng observes that the approval dialog in coding agents is narrated by the agent and proposes a mediator that renders the real action and binds the approval to the executed command~\cite{weng2026consent}. Wang gives a taxonomy of substitutions between approval and dispatch in agent runtimes and an approval token over principal, tool, arguments, and scope, which leaves open the substitutions that change no recorded field~\cite{wang2026approvallaundering}. Zhang et al.\ use the term \emph{approval laundering} for the case in which the record names the entry command and omits the transitive effects the toolchain triggers, and formalize a closure-bound approval over effect classes~\cite{zhang2026agentapproval}. Kumar reproduces post-approval substitution of the approved operation in agent frameworks~\cite{kumar2026loopjacking}, and Wu et al.\ name the general gap between what a principal authorizes and what an agent executes~\cite{wu2026aeg}. Signed approval records for Claude Code and Codex commit to the normalized action and the policy in force at enforcement time~\cite{zhou2026catp}, and a governed-execution architecture issues a one-shot approval bound to the exact bytes that will run~\cite{jing2026separation}. Security-vendor reports document approval prompts in coding agents that display a workspace path or a command while a symbolic link redirects the write to a different file~\cite{dokhanian2026ghostapproval,utevsky2026symjack}. Project configuration files in two coding agents run repository-supplied code once the project is trusted, before the first prompt~\cite{frichette2026firstprompt}. Closest in mechanism, Zhu and Wang bind a grant for an agent skill to the root of the skill's transitive dependency closure and re-resolve the closure at operation time~\cite{zhu2026closurebound}. Tiwari et al.\ bind each claim an agent makes about code to the Merkle hash of that code's dependency cone, so that the claim goes stale when the code or anything it depends on changes~\cite{tiwari2026assay}. The defenses among these works bind an approval to the executed command, to recorded dispatch fields, to a set of transitive effects, or to the closure of a skill or a code module. Our binding is to the content of the repository files that a deployment, release, or policy run is configured to read. We also measure two quantities that such a binding depends on: how often the approved object stays fixed while those files change (296 of 542 commits), and whether an agent can name them (five sets for one task).

\paragraph{Poisoned pipelines and time of check to time of use.} Indirect poisoned pipeline execution is the attack in which a protected pipeline definition is left alone and a file it references is modified~\cite{owasp_cicd_sec4}. The race between a review and a checkout in pull-request workflows is documented with worked examples and a static query~\cite{lobacevski2021pwnrequests,khan_actionstoctou,codeql_untrusted_checkout_toctou}. A deployment-branch restriction on GitHub environments was itself evaluated against a pull request's head branch while the workflow ran from the base branch, until a platform fix in December 2025~\cite{tausch2026pullrequests}. Agent-specific variants cover state that an agent validates and that changes before use~\cite{lilienthal2025mindthegap,rashidi2026balkanization}. Shipped fixes bind one script operand~\cite{openclaw2026ghsaxf99} or require re-approval when a trusted entry changes~\cite{cursor2025cve54136}. Authority-set binding carries those fixes from the named object to the files it references.

\paragraph{Supply-chain integrity.} An in-toto layout states the steps of a supply chain, who may perform them, and rules over the materials and products of each step, and verification checks recorded evidence against it~\cite{torresarias2019intoto}. An authority-set receipt has the form of the materials recorded for one step. A layout does not supply the set itself for a given entrypoint, and that set is what we audit and measure. The Update Framework~\cite{samuel2010survivable} and Sigstore~\cite{newman2022sigstore} protect the distribution and signing of artifacts, and artifact attestations record the workflow, commit, and event behind a build~\cite{github_attestations}.

\paragraph{Workflow and agent security.} Measurement of GitHub workflows finds that nearly all workflows are over-privileged and nearly all repositories run externally developed actions~\cite{koishybayev2022characterizing}, and assessment tools flag workflow misconfiguration~\cite{benedetti2022automatic}. Agentic workflow injection carries untrusted event content into agent prompts in CI~\cite{wang2026awi}, and indirect prompt injection manipulates an agent through content it reads~\cite{greshake2023indirect,zhan2025adaptive}. Those attacks corrupt the agent's decision, and a delegated-file change needs no such corruption.

\section{Ethics Considerations}

All repositories in the corpus are public and were read at pinned revisions. Every edit and every agent run took place on local copies. No change was pushed to any repository, no workflow was triggered on a hosted runner, and no cloud resource was created.

\section{Conclusion}

In the history of 20 public repositories, a digest of the approved entrypoint passes 296 of the 542 commits that change a file the approved run reads. Binding the authority set closes that gap for approvals that cannot be pinned to a commit, and a standing approval bound this way is re-requested on about one commit in ten. Two coding-agent runtimes asked to produce the binding compute the digest correctly and bind different sets from run to run: one Terraform deployment draws five sets in 16 runs, and two of those runs bind no Terraform file. The runtimes also flag a single-file task as needing more than one file in 15 of 16 runs. An approval gate should compute the digest itself, take its authority set from a source it can check, and treat the agent's set as a proposal.

\bibliographystyle{plain}
\bibliography{references}

\end{document}